# Combating COVID-19 using Generative Adversarial Networks and Artificial Intelligence for Medical Images: A Scoping Review

*Hazrat Ali, Zubair Shah**
*College of Science and Engineering Hamad Bin Khalifa University, Doha, Qatar*
*Email: haali2@hbku.edu.qa, zshah@hbku.edu.qa*

## Abstract

**Background:** Research on the diagnosis of COVID-19 using lungs images was limited by the scarcity of images data. Generative Adversarial Networks (GANs) are popular for synthesis and data augmentation. GANs have been explored for data augmentation to enhance the performance of Artificial Intelligence methods for the diagnosis of COVID-19 within lungs CT and X-Ray images. However, the role of GANs to overcome data scarcity for COVID-19 is not well understood.

**Objective:** This review presents a comprehensive study on the role of GANs in addressing the challenges related to COVID-19 data scarcity and diagnosis. It is the first review that summarizes the different GANs methods and the lungs images datasets for COVID-19. It attempts to answer the questions related to applications of GANs, popular GAN architectures, frequently used image modalities, and the availability of source code.

**Methods:** A search was conducted on five databases, namely Pubmed, IEEEXplore, ACM Digital Library, Scopus, and Google Scholar. The search was conducted between 11 October to 13 October 2021. The search was conducted using intervention keywords such as generative adversarial networks or GANs and application keywords such as COVID-19 and coronavirus. The review was performed following the guidelines of PRISMA-ScR for systematic and scoping reviews. Only those studies were included that reported GANs based methods for chest X-ray images, chest CT images, and chest ultrasound images. Any studies that used deep learning methods but did not use GANs were excluded. No restrictions were imposed on the country of publication, study design, or outcomes. Only those studies that were in English and were published from 2020 to 2022 were included. No studies before 2020 were included.

**Results:** This review included 57 full-text studies that reported the use of GANs for different applications in COVID-19 lungs images data. Most of the studies (n=42) used GANs for data augmentation to enhance the performance of AI techniques for COVID-19 diagnosis. Other popular applications of GANs were segmentation of lungs and super-resolution of the lungs images. The cycleGAN and the conditional GAN were the most commonly used architectures used in nine studies each. 29 studies used chest X-Ray images while 21 studies used CT images for the training of GANs. For majority of the studies (n=47), the experiments were done and results were reported using publicly available data. A secondary evaluation of the results by radiologists/clinicians was reported by only two studies.

**Conclusion:** Studies have shown that GANs have great potential to address the data scarcity challenge for lungs images of COVID-19. Data synthesized with GANs have been helpful to improve the training of the Convolutional Neural Network (CNN) models trained for the diagnosis of COVID-19. Besides, GANs have also contributed to enhancing the CNNs performance through the super-resolution of the images and segmentation. This review also identified key limitations of the potential transformation of GANs based methods in clinical applications.

## Introduction

### Background

In December 2019, the COVID-19 broke out and spread at an unprecedented rate, given the highly contagious nature of the virus. As a result, the World Health Organization (WHO) declared it a global pandemic in March 2020 [1]. Therefore, a response to combat the spread through speedy diagnosis became the most critical need of the time. A common method for diagnosing COVID-19 is the use of a real-time reverse transcription-polymerase chain reaction (RT-PCR) test. However, with the increasing number of cases worldwide, the health care sector was overloaded as it became challenging to cope with the requirements of the tests with the available testing facilities. Besides, research studies showed that RT-PCR may result in false negatives or fluctuating results [2]. Hence, diagnosis through computed tomography (CT) and X-Ray images of lungs may supplement the performance. Motivated by this need, alternative methods such as automatic diagnosis of COVID-19 from lungs images were explored and encouraged. In this regard, it is well understood that Artificial Intelligence (AI) techniques could help inspect chest CTs and X-rays within seconds and augment the public health care sector. The use of properly trained AI models for diagnosis of COVID-19 is promising for scaling up the capacity and accelerating the process as computers are, in general, faster than humans in computations.

Many artificial intelligence (AI) and medical imaging methods were explored to provide support in the early diagnosis of COVID-19, for example AI for COVID-19 [3-5], machine learning for COVID-19 [6], and data science for COVID-19 [7]. However, AI techniques rely on large data. For example, training a convolutional neural network (CNN) to perform classification of COVID-19 vs. normal chest X-Ray images requires training of the CNN with a large number of chest X-Ray images both for COVID-19 and normal cases. Since the diagnosis of COVID-19 requires studying of lungs CT or X-Ray images, the availability of lungs imaging data is vital to develop medical imaging methods. However, the lack of data for COVID-19 hampered the initial progress on developing these methods to combat COVID-19.

Many early attempts were made to collect image data for lungs infected with COVID-19 – specifically CT and X-Ray images either through a private collection in hospitals or through crowdsourcing using public platforms. In parallel to this, many studies explored the use of Generative Adversarial Networks (GANs) to generate synthetic image data that can improve the training of AI models to diagnose COVID-19.

GANs are a family of deep learning models that consist of two neural networks trained in an adversarial fashion [8-15]. The two neural networks, namely the generator and the discriminator, attempt to minimize their losses while maximizing the loss of the other. This training mechanism improves the overall learning task of the GAN model, particularly for generating data. GANs have recently been studied for computer vision and medical imaging tasks such as image generation, super-resolution, and segmentation [9, 10]. Given the significant potential of GANs in medical imaging, it was intuitive that many researchers were tempted to explore the use of GANs for data augmentation of imaging data for COVID-19. In addition, some researchers also used GANs for segmentation and super-resolution of lungs images. This scoping review focuses on providing a comprehensive review of the GANs based methods used to combat COVID-19. Specifically, it covers the studies where GANs have been used for lungs CT and X-Ray images to diagnose COVID-19 or to enhance the performance of CNNs for the diagnosis of COVID-19 (for example, by data augmentation or super-resolution).

## Research Problem

GANs have gained the attention of the medical imaging research community. As the COVID-19 pandemic continued to grow in 2020 and 2021, the research community faced a significant challenge due to the scarcity of medical image data on COVID-19 that can be used to train AI models (for example, CNN) to perform COVID-19 diagnosis automatically. Given the popularity of GANs for image synthesis, researchers turned to exploring the use of GANs for data augmentation of lungs radiology images. Many studies were conducted to use different variants of GANs for data augmentation of lungs CT images and lungs X-Ray images. Similarly, few studies also used GANs for the diagnosis of COVID-19 from lungs radiology images. However, to the best of our knowledge, there is no review to study the role of GANs in addressing the challenges related to COVID-19 data scarcity and diagnosis. The following research questions related to COVID-19 image data were considered for this review.

1. What were the common applications of GANs proposed for challenges related to COVID-19?
2. Which architectures of GANs are most commonly applied for data augmentation tasks related to COVID-19?
3. Which imaging modality is the popular choice for the diagnosis of COVID-19?
4. What were the most commonly used datasets of CT and X-Ray images for COVID-19?
5. What studies were conducted with open source code to reproduce the results?
6. What studies were conducted and presented to radiology experts for evaluation of the suitability towards future use in clinical applications?

The results of this review will be helpful for researchers and professionals in the medical imaging and healthcare domain who are considering using GANs methods to address challenges related to COVID-19 imaging data and to address the challenge of improving the automatic diagnosis using radiology images.

## Methods

In this work, a scoping review was conducted following the guidelines of "Preferred Reporting Items for Systematic Reviews and Meta-Analyses Extension for Scoping Reviews" (PRISMA-ScR) [16]. The methods for performing the study are described below.

### Search Strategy

#### *Search Sources*

A search was conducted between 11 October to 13 October 2021. The search was performed on the following five databases: Pubmed, IEEEXplore, ACM Digital

Library, Scopus, and Google Scholar. In the case of Google Scholar, only the first 99 results were retained as the results beyond 99 items were highly irrelevant to the scope of the study. Similarly, in the case of ACM Digital Library, the first 100 results were retained as a lack of relevancy to the study was obvious in results beyond 100.

### *Search Terms*

The search terms used in this study were chosen from the literature with guidance from experts in the field. The terms were chosen based on the intervention (for example, generative adversarial networks, GANs, cycleGANs) and the target application (COVID-19, coronavirus, corona pandemic). The exact search strings used in the search for this study are available in Appendix 1.

### Search Eligibility Criteria

This study focused on the applications of GANs in radiology images of lungs for COVID-19 used for any purpose such as data augmentation or synthesis, diagnosis, super-resolution, and prognosis. Only those studies were included that reported GANs based methods for chest x-ray images, chest CT images and chest ultrasound images. Studies that reported GANs based methods for non-lungs images were removed. Any studies that used deep learning methods but did not use GANs were also excluded. Studies reporting GANs for non-image data were also excluded. To provide a list of reliable studies, only peer-reviewed articles, conference papers and book chapters were included. Preprints, conference abstracts, short letters, and commentaries were excluded. Similarly, review articles were also excluded. No restrictions were imposed on the country of publication, study design, or outcomes.

Studies that were written in English and were published from 2020 to 2022 were included. No studies before 2020 were included.

### Study Selection

Two reviewers namely authors HA and ZS screened the titles and abstracts of the search results. Initial screening by the two reviewers was performed independently. The disagreement occurred for 9 articles only. The disagreement was resolved through mutual discussion and consensus. For measuring the disagreement, Cohen Kappa [17] was calculated to be 0.89, which shows good agreement between the two independent reviewers. Appendix 2 shows the matrix for the agreement between the two independent reviewers.

### Data Extraction

Appendix 3 shows the form for extraction of the key characteristics. The form was pilot-tested and refined in two rounds, firstly by data extraction for five studies and then by data extraction for another five studies. This refinement of the form ensured that only relevant data is extracted from the studies. The two reviewers (HA and ZS) extracted the data from the included studies, related to the *GANs-based method, applications*, and *data sets*. Any disagreement between the reviewers was resolved through mutual consensus and discussions. As the disagreements at the study selection stage were resolved through careful and lengthy discussions, the disagreement at the data extraction was only minor.

### Data Synthesis

After extraction of the data from the full text of the identified studies, a narrative approach was used to synthesize the data. The use of GANs methods was classified in terms of the application of GAN (for example, augmentation, segmentation of lungs, etc.), the type of GAN architecture, if reported (for example, conditional GAN or cycleGAN), and the modality of the imaging data for which the GAN was used (for example, CT or X-Ray imaging). Similarly, the studies were classified based on the availability of the dataset (for example, public or private), the size of the dataset (for example, the number of images in the original images and the number of images after augmentation with GAN, if applicable), and the proportion of the training and test sets as well as the type of cross-validation. The data synthesis was managed and performed using Microsoft Excel (Microsoft Corporation) workbook.

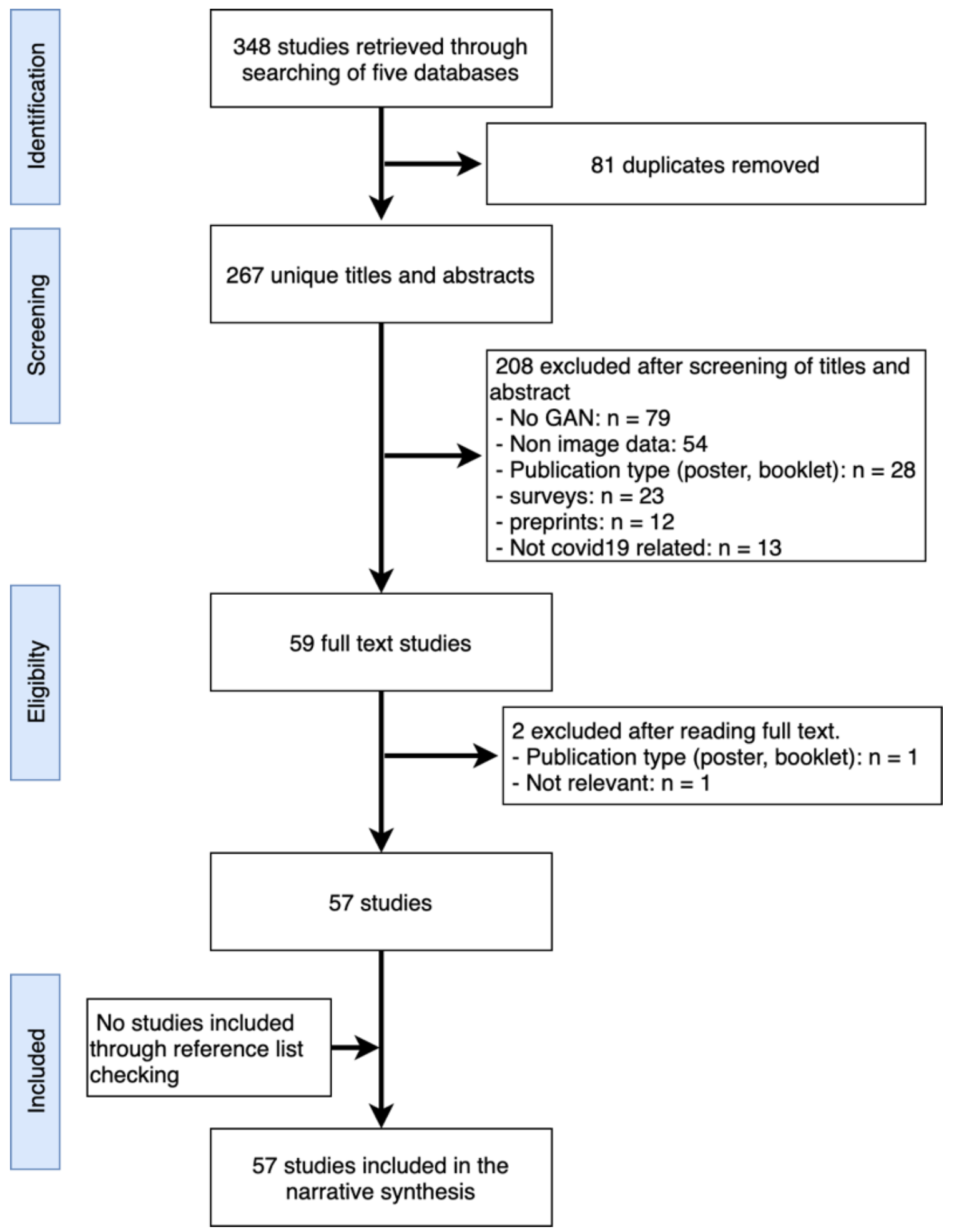

Figure 1. PRISMA-ScR flowchart for the search outcomes and selection of studies

## Results

### Search Results

From five online databases, a total of 348 studies were retrieved (see Figure 1). Out of the 348 studies, 81 duplicates were removed. The title and abstracts of the remaining 267 studies were carefully screened as per the criteria of inclusion and exclusion. The screening of the titles and abstracts resulted in the exclusion of 208 studies (see Figure 1 for reasons of exclusion). After the full-text reading of the remaining 59 studies, 02 studies were excluded following the inclusion/exclusion criteria. Finally, a total of 57 studies were included in this review. No additional studies were found through reference list checking. As per the year-wise publication, 15 out of 57 studies were published in 2020, and 41 out of 57 were published in 2021.

## Demographics of the included studies

Among the included studies (n=57), 37 studies were published articles in peer-reviewed journals, 18 studies were published in conference proceedings, and 2 studies were published as book chapters. No thesis publication was found relevant to the scope of this review. Around one-fourth of the studies (n=15) were published in the year 2020. Most of the studies were published in 2021 (n=41). The included studies were published in 14 countries. The largest number of publications were from China (n=12), followed by India (n=10). Both USA and Egypt published the same number of studies (n=6). The characteristics are summarized in (Table 1). (Figure 2) shows the demographics of the included studies along with the modality of the chest images used.

Table 1. Characteristics of the included studies. Demographics are shown for type of publication, country of publication and year of publication.

| Characteristics | | Number of studies (n) |
|---|---|---|
| Publication type | Journal | 37 |
| | Conference | 18 |
| | Book Chapter | 2 |
| Country | China | 12 |
| | India | 10 |
| | USA | 6 |
| | Egypt | 6 |
| | Canada | 4 |
| | Spain | 3 |
| | Malaysia | 2 |
| | Turkey | 2 |
| | Pakistan | 2 |
| | Vietnam | 1 |
| | Mexico | 1 |
| | South Korea | 1 |
| | Philippines | 1 |
| | Israel | 1 |
| Year of publication | 2020 | 15 |
| | 2021 | 41 |
| | 2022 | 1 |

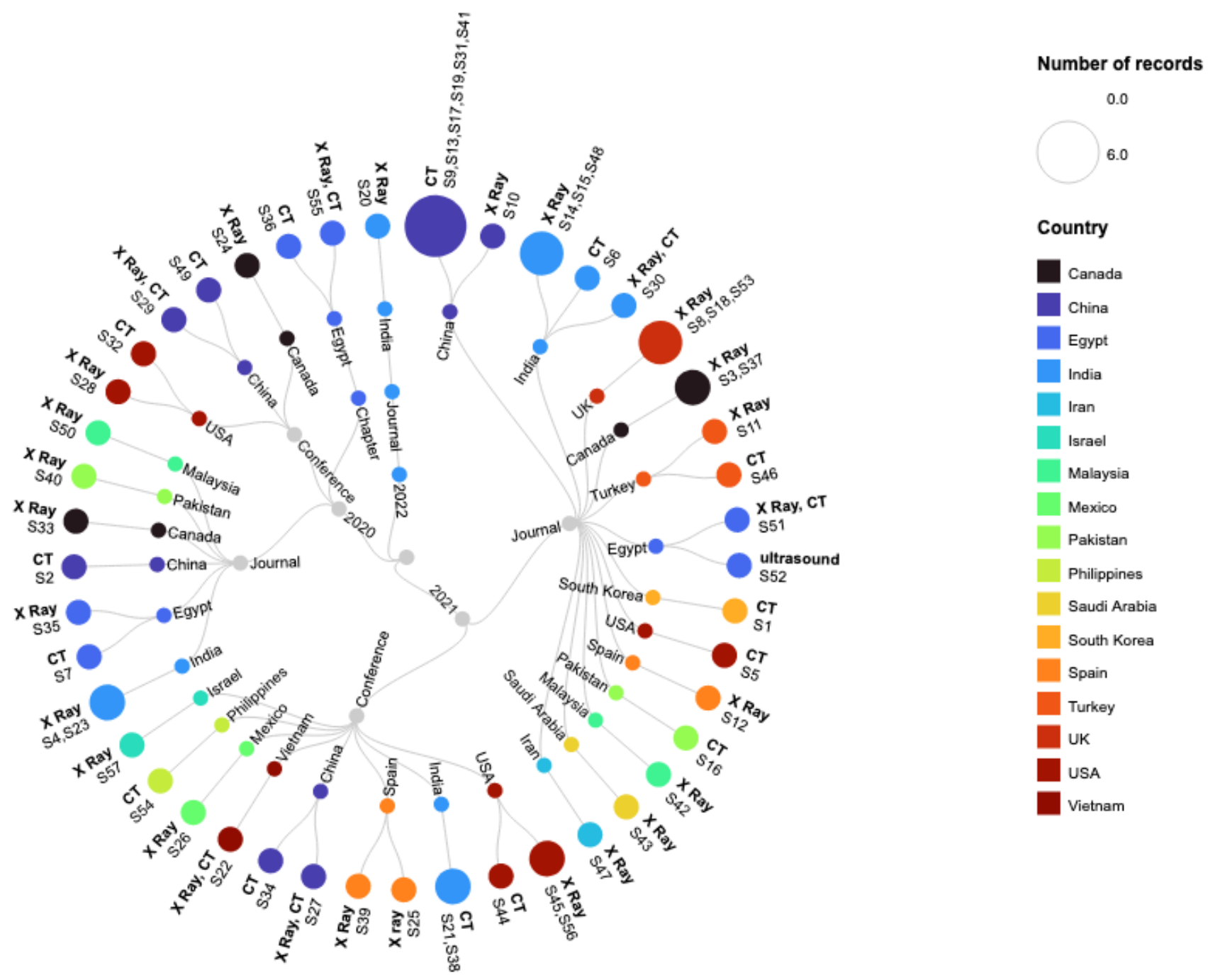

Figure 2. Characteristics of the included studies showing the publication type, country of publication, the modality of data. The number of studies is reflected by the size of terminal node. The numbers S1 to S57 refer to the included studies, as per (18) through (74) in the references.

## Application of the studies

As shown in (Table 2), the included studies have reported five different tasks being addressed: augmentation (data augmentation), diagnosis of COVID-19, prognosis, segmentation (to identify lungs region), and diagnosis of lungs disease. As the diagnosis of COVID-19 using medical imaging has been a priority since the pandemic started, 39 out of 57 studies [19-21], [23 – 33], [35 – 37], [39], [41, 42], [44], [46], [50], [52], [53], [55], [56], [58 – 60], [63 – 69], [71, 72] have reported the diagnosis of COVID-19 as the main focus of their works. Nine studies [18], [43], [45], [49], [54], [61, 62] reported data augmentation as the main task addressed in the work. In addition, one study [22] reported prognosis of COVID-19 disease, three studies reported segmentation of lungs [34], [51], [57], one study reported diagnosis of multiple lungs diseases [47].

The majority of the studies used GANs to augment the data where they reported the use of GANs to increase the dataset size. Specifically, 42 studies [18], [21], [23], [24 – 29], [31 – 36], [38 – 43], [45], [46], [48], [50], [52 – 56], [59 – 67], [71], [73, 74] used GANs based methods for data augmentation. The augmented data were then used to improve the training of different CNNs to diagnose COVID-19. Three studies [37], [51], [57] used GANs for segmentation of the lungs region within the chest radiology images. Three studies [30], [44], [68] used GANs for super-resolution to improve the quality of the images before using the images for diagnosis purposes. Five studies [20], [58], [69, 70], [72] used GANs for the diagnosis of COVID-19. Two studies [19, 47] used GANs for features extraction from images and one study [22] used GANs method for prognosis of the COVID-19 disease. The prevalent mode of imaging is the use of 2D image data and one study reported GANs based method for synthesizing 3D data [49]. (Figure 3) shows the mapping of the applications of GANs based methods for all the included studies.

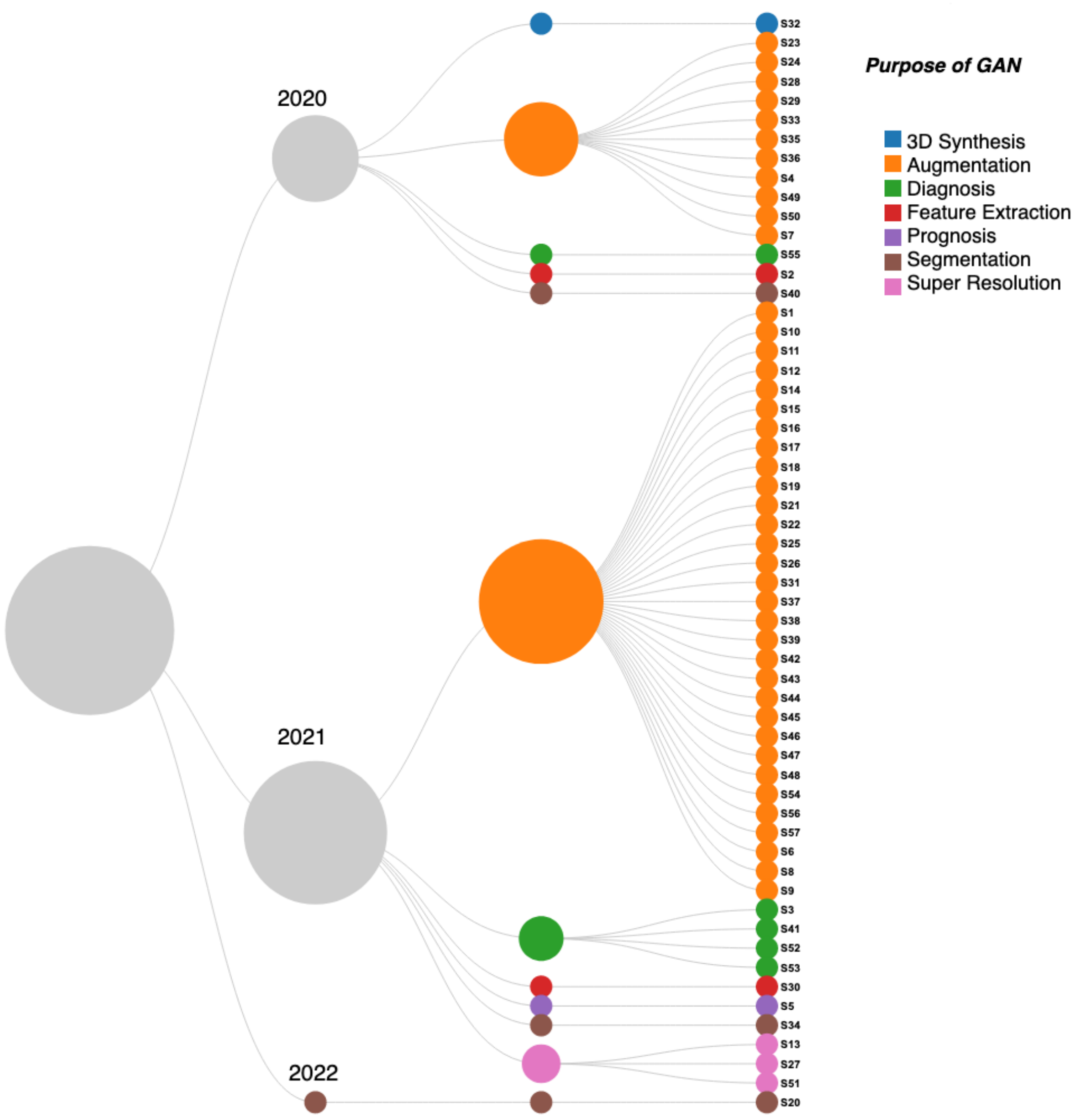

Figure 3. Major applications of GANs in the included studies. The number of publications for each application is reflected by the size of the circle in the second-last layer. The numbers S1 through S57 refer to the included studies, as per (18) through (74) in the references.

Table 2. Applications of using GANs based method and type of GANs

| | | |
|---|---|---|
| **Applications addressed in the studies** | **Applications** | **Number of studies** |
| | Diagnosis | 39 |
| | Data augmentation | 9 |
| | Segmentation + Diagnosis | 3 |
| | Segmentation | 3 |
| | Diagnosis of lungs disease | 1 |
| | Prognosis | 1 |
| | Prognosis + Diagnosis | 1 |
| **Applications of using GANs** | **Applications** | **Number of studies** |
| | Augmentation | 42 |
| | Diagnosis | 5 |
| | Super-resolution | 3 |
| | Segmentation | 3 |
| | Features extraction | 2 |
| | Prognosis | 1 |
| | 3D synthesis | 1 |
| **Type of GANs used** | **Type of GAN** | **Number of studies** |
| | GAN | 17 |
| | cycleGAN | 9 |
| | C GAN | 9 |
| | DC GAN | 4 |
| | Auxiliary Classifier GAN | 4 |
| | SR GAN | 2 |
| | 3D CGAN | 2 |
| | BiGAN | 1 |
| | Random GAN | 1 |
| | Pix2pix GAN | 1 |

Different variants have been proposed for GANs architectures since their inception. The most common type of GAN used in these studies was the cycleGAN, used in 9 studies [29], [35, 36], [42], [46], [54], [56], [70], [74]. The cycleGAN is an image translation GAN that does not require paired data to transform images from one domain to another. Other popular types of GANs were conditional GAN used by nine studies [18], [22], [24, 25], [33], [37], [41], [57], [60], Deep Convolutional GAN used by four studies [21], [38], [43], [67], and auxiliary classifier GAN used by four studies [32], [40], [55], [69]. The super-resolution GAN was used by two studies [44], [68]. One study [31] reported the use of multiple GANs namely, Wassertein

GAN, Auxiliary Classifier GAN and Deep Convolutional GAN and compared their performances for improving the quality of images.

Out of 57 studies, only ten studies [18, 19], [26, 27], [30], [34], [43], [61 – 73] reported changes to the architecture of the GAN they were using. For the rest of the studies, no major changes were reported to the architecture of the GAN.

### Characteristics of the data sets

The included studies have applied GANs on lungs radiology images of various modalities. Specifically, the use of X-Ray images dominated the studies. 29 studies [20, 21], [25], [27 – 29], [31, 32], [35], [37], [40 – 43], [45], [50], [52], [54], [56], [57], [59, 60], [62], [64, 65], [67], [70], [73, 74] used X-Ray images of lungs while 21 studies [18, 19], [22 – 24], [26], [30], [33, 34], [36], [38], [48, 49], [51], [53], [55], [58], [61], [63], [66], [71] used CT images. Six studies [39], [44], [46], [47], [68], [72] reported the use of both the X-Ray and CT images. Only one study [69] used ultrasound images for COVID-19 diagnosis which shows that the ultrasound is not a popular imaging modality for training GANs and other deep learning models for COVID-19 detection (also see Figure 4). Out of the 57 studies, most of the studies (n=47) have used images data sets that are publicly available. For ten studies, the data sets used in the work are private. (Table 3) provides a list of the various datasets used in the included studies and the public access links for the publicly available datasets. The most commonly used dataset was the COVIDx dataset available on Github and used by 26 studies.

Majority of the studies reported the size of the data set in terms of the number of images. The number of images used was greater than 10,000 in only seven studies

[20], [22], [30], [39], [63], [66], [74]. Three studies used images between 5000 and 10,000 [33], [47], [64]. The most common range for the number of images used was between 1000 to 5000 images used in 15 studies. Around one-fifth of the studies (n=11) used the number of images between 500 and 1000. In 11 studies, the number of images used was less than 500. No study reported a number of images less than 100. The maximum number of images was 84971 used by [22]. Only a few of the studies reported the number of patients for whom the data has been used. One study [26] used data for more than 1000 patients. Two studies [29], [42] used data for 500 to 1000 patients. Six studies [19], [22], [24], [30], [38], [71] used data for 100 to 500 patients. Four studies [18], [49], [66], [69] used data for less than 100 patients. The number of patients was not reported in the rest of the studies.

After augmentation using GANs, the studies have increased the number of images to several thousand with a maximum number of 21295 [54]. In six studies using GANs for data augmentation, the number of images was increased to more than 10,000. In three studies, the number of images was increased between 5000 to 10,000. In nine studies, the increased number of images was between 1000 to 5000 and in two studies, the increased number of images was between 500 and 1000. No study reported data augmentation output below 500 images.

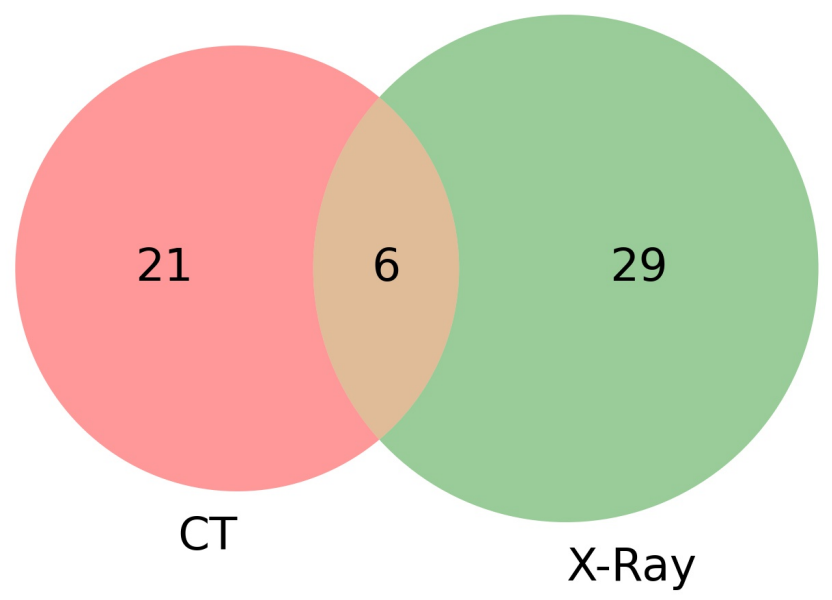

Figure 4. Venn diagram showing the number of studies using CT vs X-Ray images. Only one study reported the use of ultrasound images (not reflected here).

## Evaluation Mechanisms

Generally, the popular metrics for evaluating the diagnosis and classification performances for neural networks are accuracy, precision, recall, dice score, and area under the ROC curve. To evaluate the performance of diagnosis of COVID-19, 38 studies used accuracy along with metrics like precision, recall and dice score [21], [23 – 28], [31 – 34], [36], [38], [40], [43 – 48], [52, 53], [55], [56], [58 – 60], [63 – 72], [74]. Around one-fourth of the studies (n=18) used sensitivity and specificity. 12 studies used AUC [19], [20], [26], [30], [32], [46], [47], [48], [50], [51], [68], [74]. The numbers do not add up as many studies used more than one metric for evaluation. Besides the metrics mentioned above, only one study [22] used additional metrics, namely concordance index and relative absolute error, to evaluate prognosis and survival prediction for COVID-19 affected individuals.

Likewise, the popular metrics to assess the quality of the synthesized images are SSIM, PSNR and FID. In the included studies, six studies used the SSIM metric [18], [30], [49], [60-62], five used PSNR [18], [30], [49], [61, 62] and three used the FID metric for evaluation [18], [43], [62].

Majority of the studies (n=42) reported having the data split between independent training and test set. Few of the studies (n=6) reported 5-fold or 10-fold cross-validation for training and evaluation of the model. For almost one-sixth of the studies (n=9), the information on the cross-validation was not available.

Table 3. Resources of the data sets used in the included studies. Name is provided only if available.

| Platform (Name) | Full URL | Modality of images |
|---|---|---|
| Kaggle | https://www.kaggle.com/plameneduardo/sarscov2-ctscan-dataset | CT |
| Github | https://github.com/UCSD-AI4H/COVID-CT | CT |
| Github | https://github.com/wang-shihao/HKBU_HPML_COVID-19 | CT |
| Github (Covidx) | https://github.com/ieee8023/covid-chestxray-dataset | X-Ray, CT |
| Github | https://github.com/agchung/Actualmed-COVID-chestxray-dataset | X-Ray |
| Kaggle (Tawsif) | https://www.kaggle.com/tawsifurrahman/covid19-radiography-database | X-Ray |
| Github | https://github.com/agchung/Figure1-COVID-chestxray-dataset | X-Ray |
| Kaggle | https://www.kaggle.com/paultimothymooney/chest-xray-pneumonia | X-Ray |
| Mendeley | https://data.mendeley.com/datasets/8h65ywd2jr/3 | CT |
| Website | https://medicalsegmentation.com/covid19/ | CT |
| Kaggle (Allen Institute) | https://www.kaggle.com/allen-institute-for-ai/CORD-19-research-challenge | CT |
| Kaggle (RSNA) | https://www.kaggle.com/c/rsna-pneumonia-detection-challenge/data | X-Ray |
| Website | http://ictcf.biocuckoo.cn/HUST-19.php | CT |
| Github | https://github.com/jannisborn/covid19_ultrasound | Ultrasound |
| Kaggle | https://www.kaggle.com/bachrr/covid-chest-xray | X-Ray |
| Website (Italian Society of Medical and Interventional Radiology) | https://sirm.org/category/senza-categoria/covid-19/ | X-Ray |
| private | First Affiliated Hospital of University of Science and Technology China | CT |
| private | Massachusetts General Hospital, Brigham and Women's Hospital, | CT |
| private | Comlejo Hospitalario Universitario de A Coruna Spain | X-Ray |

## Reproducibility and Secondary Evaluation

This review also summarizes the studies for which the authors provided the implementation code. Only seven [19, 20], [34], [47, 48], [66], [70] out of the 57 studies provided links for their code. Only two studies [19, 45] reported a secondary evaluation by radiologists/doctors/experts by presenting the outcome of the results obtained by their model. One study [19] presented their results of an end-to-end diagnosis COVID-19 from CT images to three radiologists for a second opinion. One study [45] presented the synthetic X-Ray images to two radiologists for a second opinion on the quality of the generated X-Ray images.

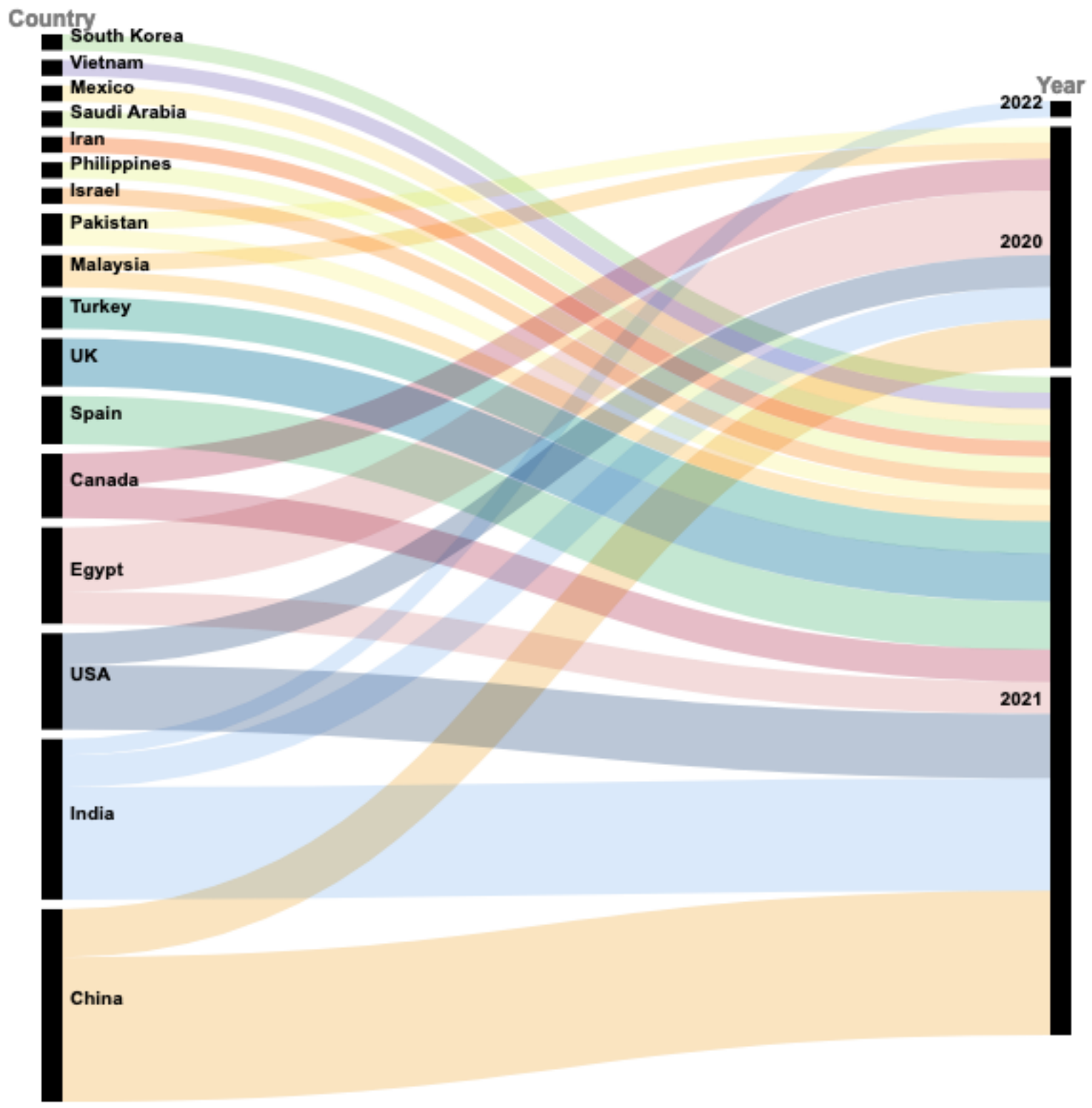

Figure 5. Mapping of correlation between publications from each country vs year of publication. Studies in 2020 are originating mostly from China, India, Egypt, Canada only. In 2021, many other countries are also contributing to the published research.

## Discussion

### Principal Results

In this review, a significant rise in the number of studies on the topic was found in the year 2021 as compared to 2020. This makes sense as the first half of the year 2020 saw only initial cases of the COVID-19 infections, and research on the use of GANs for COVID-19 had yet to gain a pace. Lung radiology images data for COVID-19 positive examples gradually became available during this period and increased only in the latter part of 2020. The highest number of studies were published from China and India (n=22). There can be two possible reasons for this. Firstly, the two countries hold the top two spots on the rankings of the world's most populous countries. Secondly, the COVID-19 pandemic started in China, hence, prompting earlier research efforts there.

Interestingly, the same number of studies (n=6) were published from USA and Egypt. The correlation mapping in (Figure 5) shows that most of the studies published in 2020 originated from China, India, Egypt, and Canada. However, in 2021, many other countries also contributed to the published research. The number of journals papers was twice of the conference papers. This is surprising as journal publications would typically require more time in article processing than conferences. It can be possible that many authors turned to journals submissions as, during the start of the pandemic, many conferences were suspended initially before moving to the online (virtual) mode.

In the majority of the included studies (n=39), the main task was to perform diagnosis of COVID-19 using lungs CT or X-Ray images. For these studies, GAN was used as a sub-module of the overall framework, and the diagnosis was performed with the help of variants of convolutional neural networks such as ResNet, VGG16, Inception-net, etc. In the included studies, GANs were used for seven different purposes: data augmentation, segmentation of lungs within chest radiology images, super-resolution of lungs images to improve the quality of the images, diagnosis of COVID-19 within the images, features extraction, prognosis studies related to COVID-19, and synthesis of 3D volumes of CT. Around 73% of the included studies used the GANs methods for data augmentation to address the data scarcity challenge of COVID-19. It is not unexpected as data augmentation is the most popular application of GANs. Only one study used the 3D variant of GAN for 3D volumes synthesis of CT. This is not surprising as the synthesis of 3D volumes using 3D GANs is computationally very expensive. The computations for the 3D synthesis of CT volumes may exceed the available resources of the GPU.

Since there are many variants of GANs, this review also looked at the most commonly used GAN architecture in the included studies. The most common choice of GAN in the included studies was the cycleGAN used in nine studies. The cycleGAN is a GAN architecture that comprises two generators and two discriminators and does not require pair-to-pair training data [11]. Hence, it was a popular choice to generate COVID-19 positive images from normal images.

This review analyzed the common imaging modality for the different applications related to COVID-19. As chest X-Ray imaging and CT scans are the most popular imaging methods for studying the infection in individuals, the studies included in this review also used these two imaging modalities. Specifically, 35 studies used X-Ray images, and 21 studies used CT images. Some of the studies (n=6) also used both the CT and X-Ray images for diagnosis by training different models or for the transformation of images from X-Ray to CT. Though ultrasound imaging is not prevalent in the clinical diagnosis of COVID-19, one study reported using ultrasound images to diagnose COVID-19 with GANs. No other modality of imaging was used by the included studies.

The majority of the included studies (n=47) used data that is available publicly on Github, Kaggle, or other publicly accessible websites. These data are acquired from multiple sources (for example, collected from more than one hospital or through crowdsourcing) which makes them more diverse and hence more useful for training of GANs models. Similarly, it is hoped that the use of publicly accessible data will also encourage other researchers to conduct experiments on the data sets. The rise of publications in 2021 can also be linked to the availability of publicly available data sets that continued to rise as the number of COVID-19 infected cases continued to grow. A few of the included studies (n=10) used private or proprietary data sets, and hence, the details about those data sets are only limited to what has been described in the corresponding studies.

Only 13 studies provided information on the number of individuals whose data was used in the included studies. Amongst these, only one study [26] used data for more than 1000 individuals, and two studies [29], [42] used data for more than 500 individuals. The remaining ten studies used data for less than 500 individuals. Given the size of the population infected with the COVID-19 (418+ million as of writing this, reported from John Hopkins University Coronavirus Resource Center[1]), the need for experiments with much more extensive data is obvious. As a result of having more data, learning inherent features within the radiology images by using GANs will become more generalized with training on larger data. There is still more need to contribute to publicly accessible data.

### Practical and Research Implications

This review presented the different studies that used GANs for various applications in COVID-19. Data augmentation of COVID-19 images data was the most common application in the included studies. The augmented data can significantly improve the training of AI methods, particularly deep learning methods used for COVID-19 diagnosis. This review found that for most of the studies, the current CT and X-Ray images data (even if smaller in size) is already available through publicly accessible links on Github, Kaggle, or institutional websites. This should encourage more researchers to build upon the available data sets and train more variants of deep learning and GANs methods to speed up the research progress on COVID-19. Similarly, researchers can also add to the existing dataset on Github by uploading their data to the current data repositories. An example of crowdsourcing of data is the COVIDx images repository for lungs X-ray images (see Table 3).

This review identified that the code to reproduce the results was not available for the majority of the studies. Only seven of the included studies provided a public link to the code. Availability of a public repository to reproduce the results for diagnosis or augmented data can help in advancing the research as well as increase the trust and reliance on the reported results in terms of the quality of the generated images or the accuracy reports for the diagnosis. Besides, the reproducibility by this code is not assessed by this review as it was beyond the scope of this review. Careful and responsible studies are needed to make an assessment of the published methods for transformation into clinical applications.

[1] https://coronavirus.jhu.edu/

The majority of the included studies (n=43) did not provide information on the number of patients, although they did mention the number of images used in the experiments. So, it is unclear that how many images were used per individual. Hence, the lack of information limits the ability of the readers to evaluate the performance in the context of the number of patients. Moreover, for public data set with crowd-sourced contributions, it is challenging to trace back the number of images to the number of individuals.

Validation of the performance of GANs in terms of the quality/usability of the generated images has a significant role in promoting the acceptability of the methods. In the included studies, only two studies reported that the results were presented to radiologists/clinicians for a secondary validation. For one study on the synthesis of X-Ray images, the radiologists agreed that the quality of the X-Rays has improved but falls short of diagnostic quality for use in clinics [45]. While using GANs methods in COVID-19 is tempting for many researchers, the lack of evaluation by radiologists or using GANs based methods without radiologists and clinicians in the loop will hinder the acceptability of these methods for clinical applications. Besides, it is beyond the scope of this review to evaluate a study based on reporting of secondary evaluation by the radiologists, though a secondary assessment by the radiologists would have added value to the studies and increased their acceptability. The lack of details related to the individuals whose COVID-19 data were used in these studies may also hinder their acceptance for transformation into clinical applications. The training of GANs is usually computationally demanding, requiring GPUs. More edge computing-based implementations are needed for clinical applications to make these models compatible for implementation on low-power devices. This will increase the acceptability of these methods in clinical devices.

## Strengths and Limitations

### Strengths

Though several reviews can be found on the applications of AI techniques in COVID-19, no review was found that focused on the potential of GANs based methods to combat COVID-19. Compared to other reviews [3, 4], [6, 7] where the scope is too broad as they attempted to cover many different AI models, this review provides a comprehensive analysis of the GANs based approaches used primarily on lungs CT and X-Ray images. Similarly, many reviews cover the applications of GANs in medical imaging [10], [12-15]; their applications in lungs images for COVID-19 have not been reviewed before. So, this review may be considered the first comprehensive review that covers all the GANs methods used for COVID-19 imaging data for different applications in general and data augmentation in particular. Thus, it is helpful for the readers to understand how GANs based approaches were used to address the problem of data scarcity and how the synthetic data (generated by GANs) was used to improve the performance of CNNs for COVID-19. This review provided a thorough list of the various publicly available datasets of lungs CT, lungs X-Ray, and lungs ultrasound images, along with the public URL. Hence, this can serve as a single point of contact for the readers to explore these data set resources and use them in their research work. This review is consistent with the guidelines of PRISMA-ScR for scientific reviews [16].

### Limitations

This review included studies from five databases: Pubmed, IEEEXplore, ACM Digital Library, Scopus, and Google Scholar. Hence, it is possible that literature might have been left out if it is not indexed in these libraries. However, given the coverage by these popular databases, the included studies form a comprehensive representation of the applications of GANs in COVID-19. The review, for practical reasons, included studies published in English only and did not include studies in other languages. Since the scope of this review was limited to lungs images only, the potential of GANs for other types of medical data such as electronic health records, textual data, and audio data (recordings of coughing) is not covered in this review. The results and interpretations presented in this review are derived from the available information in the included studies. Since different studies may have variations and even missing details in their reporting of the dataset, the training and test sets, the validation mechanism, a direct comparison of the results might not be possible. Inconsistent information on the number of images, the training mechanism for GANs, and the selection of test set examples may have affected the findings of this review. In addition, by modern standards of training deep learning models, the size of data reported in most included studies is too small. So, the results reported in the studies in terms of diagnosis accuracy may not generalize well. The findings and the discussions of this review are mainly based on the authors' understanding of GANs (and other AI methods) and do not necessarily reflect the comments and feedback of the doctors and clinicians.

## Conclusion

This scoping review provided a comprehensive review of 57 studies on the use of GANs for COVID-19 lungs images data. Similar to other deep learning and AI methods, GANs have demonstrated outstanding potential in research on addressing COVID-19 diagnosis. However, the most significant application of GANs has been the data augmentation by generating synthetic chest CT or X-Ray images data from the existing limited size data as the synthetic data showed a direct bearing on the enhancement of the diagnosis. Although GANs based methods have demonstrated great potential, their adoption in COVID-19 research is still in a stage of infancy. Notably, the transformation of GANs based methods into clinical applications is still limited due to the limitations in the validation of the results, the generalization of the results, the lack of feedback from radiologists, and the limited explainability offered by these methods. Nevertheless, GANs based methods can assist in the performance enhancement of COVID-19 diagnosis even though they should not be used as independent tools. Besides, more research and advancements are needed towards the explainability and clinical transformations of these methods. This will pave the way for a broader acceptance of GANs based methods in COVID-19 applications.

## Acknowledgements

H.A. contributed to the conception, design, literature search, data selection, data synthesis, data extraction, drafting. Z.S. contributed to the design, data selection, data synthesis and critical revision of the manuscript. All authors gave their final approval and accepted accountability for all aspects of the work.

## Conflicts of Interest

None declared.

## Abbreviations

| Abbreviation | Full name |
|---|---|
| **ACM** | Association of Computing Machinery |
| **C GAN** | Conditional GAN |
| **COVID** | Coronavirus disease |
| **CT** | Computed Tomography |
| **CNN** | Convolutional Neural Network |
| **DC GAN** | Deep Convolutional GAN |
| **FID** | Frechet Inception Distance |
| **GAN** | Generative Adversarial Networks |
| **GPU** | GraphicsProcessing Units |
| **IEEE** | The Institute of Electrical and Electronics Engineers |
| **PSNR** | Peak Signal-to-Noise Ratio |
| **RT-PCR** | Reverse transcription-polymerase chain reaction |
| **SR GAN** | Super-resolution GAN |
| **SSIM** | Structural Similarity Index Measure |
| **WHO** | World Health Organization |